\documentclass[twocolumn,times,tighten,twocolappendix,resetfootnote,longbib]{aastex701}
\hypersetup{linkcolor=blue,citecolor=blue,filecolor=blue,urlcolor=magenta}

\usepackage{amsmath}

\begin{document}

\title{The dust-rich, gas-depleted protosolar disk as the birthplace of chondrules}

\author[orcid=0000-0003-0947-9962,sname='Arakawa']{Sota Arakawa}
\altaffiliation{These authors contributed equally to this work.}
\affiliation{Center for Mathematical Science and Advanced Technology, REIS, Japan Agency for Marine-Earth Science and Technology, \\Yokohama, 236-0001, Japan}
\email[show]{arakawas@jamstec.go.jp}  

\author[orcid=0000-0001-9659-658X,sname='Tanaka']{Hidekazu Tanaka}
\altaffiliation{These authors contributed equally to this work.}
\affiliation{Astronomical Institute, Tohoku University, Sendai, 980-8578, Japan}
\email{hidekazu@astr.tohoku.ac.jp }

\author[orcid=0000-0002-5299-556X,sname='Kadono']{Toshihiko Kadono}
\affiliation{Department of Basic Sciences, University of Occupational and Environmental Health, Kitakyushu, 807-8555, Japan}
\email{kadono@med.uoeh-u.ac.jp}

\author[orcid=0000-0002-5934-5076,sname='Ushikubo']{Takayuki Ushikubo}
\affiliation{Kochi Institute for Core Sample Research, EMS, Japan Agency for Marine-Earth Science and Technology, Nankoku, 783-8502, Japan}
\affiliation{Marine Core Research Institute, Kochi University, Nankoku, 783-8502, Japan}
\email{ushikubot@jamstec.go.jp}

\author[orcid=0000-0002-2959-0302,sname='Nagasawa']{Makiko Nagasawa}
\affiliation{Department of Physics, School of Medicine, Kurume University, Kurume, 830-0011, Japan}
\email{nagasawa_makiko@kurume-u.ac.jp}

\author[orcid=0000-0001-8808-2132,sname='Kobayashi']{Hiroshi Kobayashi}
\affiliation{Department of Physics, Nagoya University, Nagoya, 464-8602, Japan}
\email{kobayashi.hiroshi.g3@f.mail.nagoya-u.ac.jp}

\begin{abstract}

Chondrules are the primary components of primitive meteorites known as chondrites, and understanding their formation and accumulation is essential for elucidating the history of planet formation in the Solar System.
Although a variety of chondrule formation mechanisms have been proposed, it remains challenging to satisfy the key constraints on chondrule abundance, formation timing, and mineralogical and chemical characteristics within a single model.
In particular, the planetesimal bow-shock model, once considered one of the leading candidates, now faces a fundamental difficulty: Jupiter’s formation likely depleted gas in the protosolar disk, potentially lowering the gas density below that required for efficient chondrule formation by planetesimal bow shocks.
Here we propose an alternative mechanism that can occur in a gas-depleted environment: heavy bombardment of eccentric planetesimals by debris dust.
After Jupiter formed in the protosolar disk, the region interior to its orbit became gas-depleted, leading to the formation of a geometrically thin debris-dust layer.
When planetesimals enter the dust layer at high speed, large quantities of molten silicate droplets are produced.
These droplets cool and solidify into chondrules and are reincorporated into the dust layer.
Using analytical calculations, we find that our model can potentially explain the abundance, formation timing, and mineralogical and chemical characteristics of chondrules.
This study links the formation of Jupiter and the accompanying evolution of the protosolar disk to the origin of terrestrial planets, asteroids, and meteorites, thereby offering a new framework for the formation of the Solar System.

\end{abstract}




\section{Introduction}

Chondrules are 0.1--1-mm-sized spherical igneous grains.
The volume fraction of chondrules in ordinary chondrites is $\sim$ 60--80\% \citep[e.g.,][]{2007AREPS..35..577S}.
Most chondrules in ordinary chondrites formed at $\sim$ 2~Myr after Ca--Al-rich inclusions (CAIs), which are the first condensates in the solar protoplanetary disk, within a relatively short interval of less than 1~Myr \citep[e.g.,][]{2021GeCoA.293..103S, 2022GeCoA.324..312S}.
There is broad consensus that some chondrules experienced multiple heating events, as evidenced by their petrological and isotopic characteristics \citep[e.g.,][]{2007E&PSL.257..274R, 2018crpd.book..192T, 2021ApJ...910...70V, 2024SSRv..220...69M}.
Their igneous textures indicate rapid crystallization from a molten state, with inferred timescales of seconds to hours \citep[e.g.,][]{2018crpd.book...57J, 2025SciA...11.1187M}.
The abundances of volatile elements and the redox state of chondrule olivines suggest that chondrules formed in highly dust-enriched environments \citep[e.g.,][]{2008Sci...320.1617A, 2018crpd.book..192T}.
Taken together, these observations suggest that chondrules formed through localized, high-energy events in the early Solar System.
A variety of chondrule formation mechanisms have been proposed to date \citep[e.g.,][]{2011E&PSL.308..369A, 2013ApJ...776..101B, 2015Natur.517..339J, 2023ApJ...947...15K, 2025NatSR..1530919S, 2025PSJ.....6..108S}.
However, it remains challenging to satisfy the key constraints on chondrule abundance, formation timing, and mineralogical and chemical characteristics within a single model.

Planetesimals with large orbital eccentricities, excited by gravitational perturbations from the giant planets (Jupiter and Saturn), have been regarded as promising sources of high-energy events capable of producing large amounts of molten rock \citep[e.g.,][]{1998Sci...279..681W}.
Mean-motion resonances with Jupiter can excite the eccentricities of planetesimals in the vicinity of the asteroid belt ($\sim$ 2--3~au from the Sun), and the resulting random velocities can reach $\sim$ 5--10~km/s \citep[e.g.,][]{1998Sci...279..681W, 2002CeMDA..82..225M, 2014ApJ...794L...7N, 2019ApJ...871..110N}.
Under the ideal assumption that kinetic energy is converted into heat with 100\% efficiency, the instantaneous deceleration of rocky grains by $\sim$ 2~km/s near a planetesimal would be sufficient to raise their temperatures by more than 1000~K and melt them.
In addition, the cooling timescale of molten rocky grains is expected to be short (seconds to minutes), because the spatial scale of the heating event is comparable to the planetesimal radius.
Therefore, eccentric planetesimals in the solar protoplanetary disk have the potential to produce chondrule-like igneous objects.

Chondrule formation mechanisms involving eccentric planetesimals have mainly been discussed in terms of two types of models: planetesimal bow shocks \citep[e.g.,][]{2013ApJ...776..101B} and planetesimal collisions \citep[e.g.,][]{2011E&PSL.308..369A, 2015Natur.517..339J}.
In the planetesimal bow-shock model, planetesimals on high-eccentricity orbits move supersonically through the gas of the solar protoplanetary disk, generating bow shocks, and chondrule precursors that are dynamically coupled to the disk gas are melted in the post-shock region.
Gas drag in the post-shock region must be sufficiently strong for chondrules to form, and radiative hydrodynamic simulations have shown that, for planetesimals with radii smaller than 1000 km, a disk gas density of at least $10^{-9}~{\rm g}~{\rm cm}^{-3}$ is required \citep[e.g.,][]{2004M&PS...39.1809C, 2005ASPC..341..873H}.
However, the gas density required by this model is more than an order of magnitude higher than that at 2--3~au in the minimum-mass solar nebula \citep{1977Ap&SS..51..153W, 1981PThPS..70...35H}, and such a gas density cannot be achieved unless an exceptionally massive disk is assumed \citep[e.g.,][]{2007ApJ...671..878D}.
Moreover, the presence of Jupiter is necessary to excite planetesimals onto high-eccentricity orbits, but once gas accretion onto proto-Jupiter is taken into account, the gas density at 2--3~au can become one or two orders of magnitude lower than it would be in the absence of Jupiter \citep[e.g.,][]{2016ApJ...823...48T, 2020ApJ...891..143T}.
The gas in the region interior to Jupiter's orbit could become almost completely depleted when gas dispersal by disk winds is taken into account \citep[e.g.,][]{2020MNRAS.492.3849K}.
In summary, in the solar protoplanetary disk after Jupiter's formation, planetesimal bow shocks are unlikely to have served as a major chondrule-forming mechanism.

Planetesimal collision models have been intensively investigated in recent years, and several scenarios that differ in pre-impact thermal state and composition have been proposed \citep[e.g.,][]{2012M&PS...47.2170S, 2014ApJ...794...91D, 2018Icar..302...27L, 2022Icar..38415110C, 2025NatSR..1530919S, 2025PSJ.....6..108S}.
The most important difficulty with the planetesimal collision model is that it cannot readily account for chondrules that experienced multiple heating events \citep[e.g.,][]{2008GeCoA..72.5530R}.
In this scenario, molten rocky grains produced by planetesimal collisions are dispersed into the disk and then solidify \citep[e.g.,][]{2021MNRAS.503.3297C}, but each grain is expected to record only a single heating event.
This contrasts with bow-shock models, in which chondrules can, in principle, undergo multiple heating events while orbiting within the disk.
It should be noted that not all chondrules preserve evidence for multiple heating events.
Thus, we do not rule out the possibility that some chondrules formed through planetesimal collisions.
Indeed, chondrules in CB and CH carbonaceous chondrites are widely considered to have formed in impact-generated gas--melt plumes produced by planetesimal collisions \citep[e.g.,][]{2022M&PS...57..352K}.

It is reasonable to assume that multiple chondrule-forming mechanisms may have operated in the early Solar System.
In this study, we propose an additional plausible chondrule formation mechanism that is distinct from planetesimal collisions and is particularly relevant to chondrules in ordinary chondrites.
In Section~\ref{sec:overview}, we provide an overview of our scenario.
We then assess the viability of the proposed scenario through order-of-magnitude calculations of the chondrule production rate in Section~\ref{sec:rate}.
We also evaluate whether the proposed scenario can satisfy the thermal conditions required for producing molten chondrules in Section~\ref{sec:heat}.
The chronological aspects of chondrule formation are discussed in Section~\ref{sec:Jupiter}.
This study offers new insights into chondrule formation and its connection to the evolution of the protosolar disk.

\section{Overview}
\label{sec:overview}

Geometrically thin, dusty, and gas-depleted circumstellar disks, known as debris disks, are frequently observed around young stars \citep[e.g.,][]{2018ARA&A..56..541H}.
We note that debris disks are not necessarily cold, distant disks located at several tens to hundreds of au from their host stars.
A large number of warm debris disks with dust temperatures of $\gtrsim 150~{\rm K}$ have been identified \citep[e.g.,][]{2024AJ....167..275M}, and debris-disk candidates have also been reported around stars as young as a few Myr old \citep[e.g.,][]{2009ApJ...705.1646C, 2009AJ....138..703C}.
In this study, we propose a mechanism in which eccentric planetesimals penetrate a dusty disk at velocities of several km/s and experience intense heating due to heavy bombardment by debris dust.
This process is expected to be inevitable and ubiquitous after Jupiter's formation, when most of the gas in the inner protosolar disk had dissipated.
The kinetic energy of the colliding dust particles can efficiently heat planetesimal surfaces, producing molten rocky ejecta that solidify into chondrules.
This model can also account for other key constraints on chondrule formation, including repeated heating events in a dust-enriched environment \citep[e.g.,][]{2007E&PSL.257..274R, 2008Sci...320.1617A, 2018crpd.book..192T}.

\begin{figure}[]
\centering
\includegraphics[width = \columnwidth]{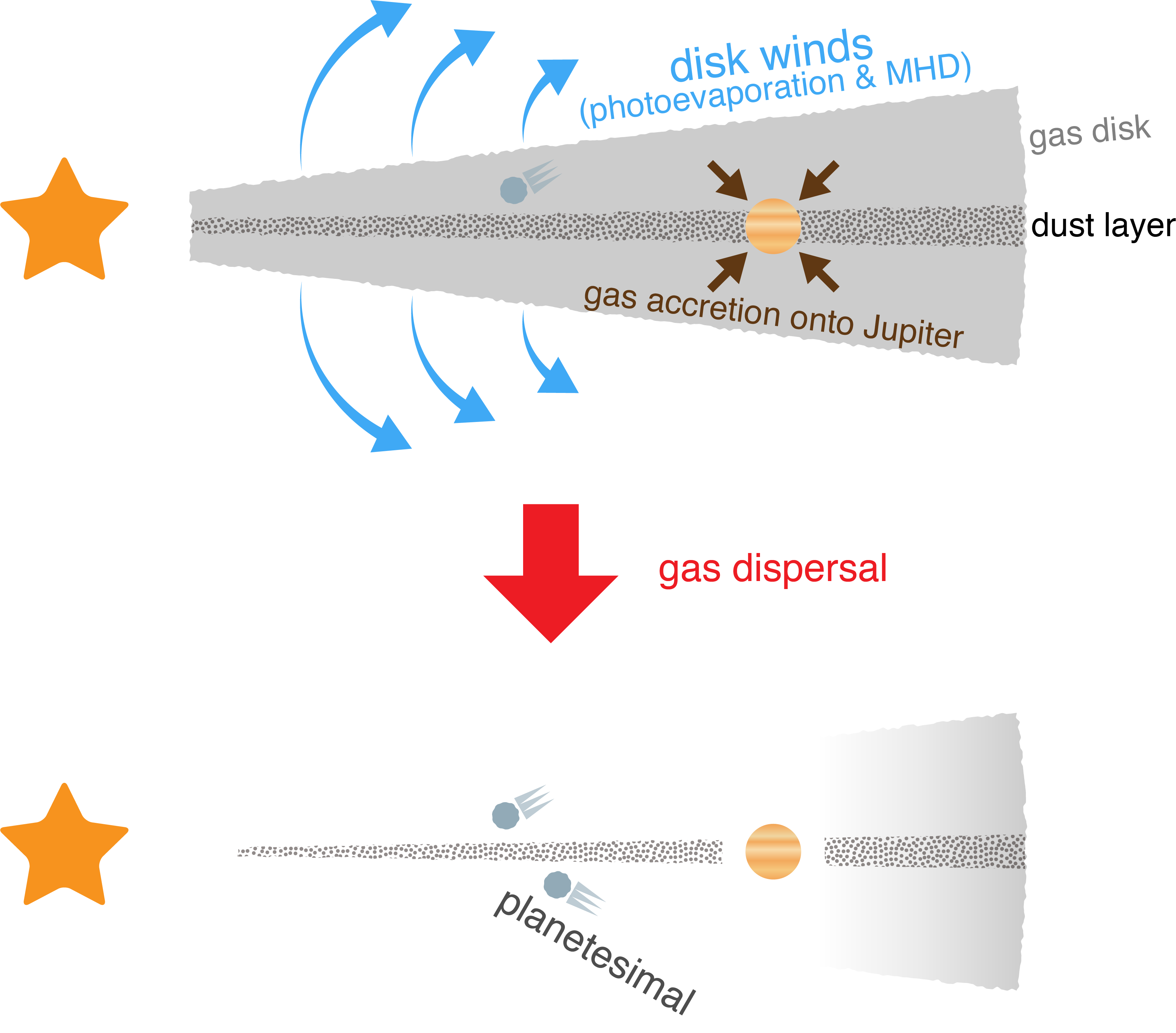}
\caption{
Schematic illustration of the inner protosolar disk evolution interior to Jupiter's orbit after the formation of proto-Jupiter.
Proto-Jupiter and planetesimals formed within the gaseous disk.
Gas accretion onto proto-Jupiter reduces the gas flow across Jupiter’s orbit, causing depletion of disk gas near and interior to the orbit.
Disk winds driven by photoevaporation and MHD processes further promote this depletion.
Depletion of disk gas could promote additional planetesimal formation, leading to the formation of an inner debris disk populated by planetesimals after gas dispersal.
}
\label{fig:1}
\end{figure}

Figure~\ref{fig:1} schematically illustrates the evolution of the inner protosolar disk interior to Jupiter's orbit after the formation of proto-Jupiter.
Astronomical observations of extrasolar protoplanetary disks have revealed the presence of 0.1--1 mm-sized dust particles in the disk midplane \citep[e.g.,][]{2020ARA&A..58..483A}.
Some disks with ages of a few Myr host planets \citep[e.g.,][]{2019NatAs...3..749H, 2024Natur.635..574B}, implying that planetesimals must also have formed on similar timescales.
Nucleosynthetic anomalies in meteorites exhibit a clear dichotomy \citep[e.g.,][]{2011E&PSL.311...93W, 2019ApJ...883...62Y, 2020SSRv..216...55K}, which may have been caused by a gap in the gaseous protoplanetary disk carved by proto-Jupiter \citep[e.g.,][]{2017PNAS..114.6712K}.
Proto-Jupiter acquired its current mass by accreting gas from the protoplanetary disk.
This accretion reduces the gas flow across Jupiter's orbit, leading to depletion of disk gas near and interior to Jupiter's orbit \citep[e.g.,][]{2016ApJ...823...48T, 2020ApJ...891..143T}.
Disk winds driven by photoevaporation and magnetohydrodynamic (MHD) processes further enhance this depletion \citep[e.g.,][]{2017RSOS....470114E, 2020MNRAS.492.3849K}.
The resulting gas depletion could increase the dust-to-gas mass ratio and thereby promote additional planetesimal formation \citep[e.g.,][]{2017ApJ...839...16C}, ultimately leading to the formation of an inner debris disk populated by planetesimals after gas dispersal.
The debris present before chondrule formation could have originated from both (i) fragments produced by planetesimal collisions and (ii) dust aggregates that grew in the protosolar disk.
Refractory condensates formed in the protosolar disk, such as CAIs, would also have been present as part of this primordial debris.

\subsection{Energetics of the Debris-dust Layer}

Here, we estimate the geometry of the gas-free debris-dust layer based on energetic considerations.
The thickness of the layer, $H_{\rm d}$, is determined by the random velocity of dust particles, $v_{\rm d}$, as $H_{\rm d} \sim v_{\rm d} / \Omega_{\rm K}$, where $\Omega_{\rm K}$ is the Keplerian angular velocity.
We define the dust surface density at heliocentric distance $r$ as $\Sigma_{\rm d}$, and the vertically averaged dust density is then given by $\rho_{\rm d} = \Sigma_{\rm d} / H_{\rm d}$.

Mutual collisions between dust particles dissipate kinetic energy and reduce $v_{\rm d}$.
Assuming that the typical radius and material density of dust particles are $R_{\rm d} = 0.1~{\rm mm}$ and $\rho_{\rm m} = 3~{\rm g}~{\rm cm}^{-3}$, respectively, the mass of a single dust particle is $m_{\rm d} = {( 4 \pi / 3 )} \rho_{\rm m} {R_{\rm d}}^{3} = 1 \times 10^{-5}~{\rm g}$.
Its geometric cross section is $A_{\rm d} = \pi {R_{\rm d}}^{2} = 3 \times 10^{-4}~{\rm cm}^{2}$.
The collision frequency of a dust particle, $f_{\rm d}$, is estimated as $f_{\rm d} \sim \rho_{\rm d} v_{\rm d} A_{\rm d} / m_{\rm d} \sim {( \Sigma_{\rm d} \Omega_{\rm K} )} / {( \rho_{\rm m} R_{\rm d} )}$.

In contrast, collisions between dust particles and eccentric planetesimals inject kinetic energy into the debris disk.
Here, we consider the motion of an eccentric planetesimal in a frame co-rotating with the debris disk at heliocentric distance $r$.
We denote the surface density of eccentric planetesimals by $\Sigma_{\rm pl}$ and their random velocity relative to Keplerian motion by $v_{\rm pl}$.
When a planetesimal's orbit is perturbed by Jupiter, its orbital inclination $i_{\rm pl}$ is generally smaller than its eccentricity $e_{\rm pl}$ \citep[e.g.,][]{2002CeMDA..82..225M}, and thus we approximate $v_{\rm pl} \sim e_{\rm pl} v_{\rm K}$, where $v_{\rm K} = r \Omega_{\rm K} = 30~{( r / 1~{\rm au} )}^{- 1/2}~{\rm km}~{\rm s}^{-1}$ is the Keplerian velocity.
Assuming that the internal density of planetesimals is equal to $\rho_{\rm m}$, the mass of a single planetesimal is $m_{\rm pl} = {( 4 \pi / 3 )} \rho_{\rm m} {R_{\rm pl}}^{3}$, where $R_{\rm pl}$ denotes the typical radius of planetesimals.
When $i_{\rm pl} \gg H_{\rm d} / r$, the orbital plane of a planetesimal no longer lies entirely within the debris-dust layer.
In this case, the planetesimal passes through a volume $V \approx \pi {R_{\rm pl}}^{2} L_{\rm cross}$ of the debris-dust layer per orbit, where $L_{\rm cross} \sim H_{\rm d} \times e_{\rm pl} / i_{\rm pl}$ is the crossing length.
The collision frequency of a dust particle with eccentric planetesimals, $f_{\rm pl}$, is then estimated as $f_{\rm pl} \sim {( V \Sigma_{\rm pl} \Omega_{\rm K} )} / {( H_{\rm d} m_{\rm pl} )} \sim {( \Sigma_{\rm pl} \Omega_{\rm K} )} / {( \rho_{\rm m} R_{\rm pl} )} \times {( e_{\rm pl} / i_{\rm pl} )}$.
We adopt fiducial parameters of $r = 2~{\rm au}$, $R_{\rm pl} = 5~{\rm km}$, $e_{\rm pl} = 0.25$, $i_{\rm pl} = 0.025~{\rm rad}$, $\Sigma_{\rm d} = 10~{\rm g}~{\rm cm}^{-2}$, and $\Sigma_{\rm pl} = 1~{\rm g}~{\rm cm}^{-2}$.
Under these conditions, the heating timescale of the debris-dust layer by eccentric planetesimals is
\begin{equation}
\label{eq:tau_pl}
\tau_{\rm pl} = {f_{\rm pl}}^{-1} \sim 1 \times 10^{-1}~{\rm Myr},
\end{equation}
and the random velocity of the planetesimals is $v_{\rm pl} \sim 5~{\rm km}~{\rm s}^{-1}$ (see Appendix~\ref{app:v_pl}).

As mutual collisions between dust particles are inelastic, the energy dissipated per collision is $e_{-} \approx {( 1 / 2 )} m_{\rm d} {v_{\rm d}}^{2}$.
Similarly, the energy gained by a single dust particle per collision with a planetesimal is $e_{+} \approx {( 1 / 2 )} m_{\rm d} {v_{\rm pl}}^{2}$.
By solving the energy balance equation, $f_{\rm d} e_{-} = f_{\rm pl} e_{+}$, we obtain the equilibrium value of $v_{\rm d}$ in the debris disk: $v_{\rm d} \sim {\left( {\Sigma_{\rm pl} R_{\rm d} e_{\rm pl}} / {\Sigma_{\rm d} R_{\rm pl} i_{\rm pl}} \right)}^{1/2} v_{\rm pl} \sim 7 \times 10^{-1}~{\rm m}~{\rm s}^{-1}$.
The thickness and density of the debris disk, $H_{\rm d}$ and $\rho_{\rm d}$, are then given by
\begin{align}
H_{\rm d}    & \sim \frac{v_{\rm d}}{\Omega_{\rm K}} \sim 1 \times 10^{4}~{\rm km},                  \label{eq:H_d} \\
\rho_{\rm d} & \sim \frac{\Sigma_{\rm d}}{H_{\rm d}} \sim 1 \times 10^{-8}~{\rm g}~{\rm cm}^{-3}.    \label{eq:rho_d}
\end{align}
In this estimate, we assume that the mass flux of debris dust impacting a planetesimal is comparable to that of debris dust ejected by collisions, and we neglect the temporal evolution of $\Sigma_{\rm d}$ and $\Sigma_{\rm pl}$.
The validity of this assumption is examined later (see Section \ref{sec:rate}).

We also briefly discuss the long-term evolution of a planetesimal-hosting debris disk.
Eccentric planetesimals at around 2--3~au gradually migrate inward through their interaction with the debris disk \citep[e.g.,][]{2019ApJ...871..110N}, leading to a decrease in $\Sigma_{\rm pl}$ over time.
According to Equation~\eqref{eq:rho_d}, the depletion of eccentric planetesimals results in an increase in $\rho_{\rm d}$, and once $\rho_{\rm d}$ reaches the Roche density, $\rho_{\rm R} \approx 5 \times 10^{-8}~{( r / 2~{\rm au} )}^{-3}~{\rm g}~{\rm cm}^{-3}$ \citep[e.g.,][]{2002ApJ...580..494Y}, part of the debris dust collapses into planetesimals via gravitational instability \citep[e.g.,][]{2010AREPS..38..493C}.
This process works as a buffer that maintains $\rho_{\rm d}$ below $\rho_{\rm R}$, and the formation of chondrite parent bodies may also be associated with this mechanism.
Additionally, the accumulation of inward-migrating planetesimals near 1~au may provide favorable conditions for the formation of the Solar System's terrestrial planets from a narrow planetesimal ring \citep[e.g.,][]{2009ApJ...703.1131H, 2016AJ....152...68W}.

\subsection{Chondrule Formation via Heavy Bombardment of Planetesimals by Debris Dust}

\begin{figure}[]
\centering
\includegraphics[width = \columnwidth]{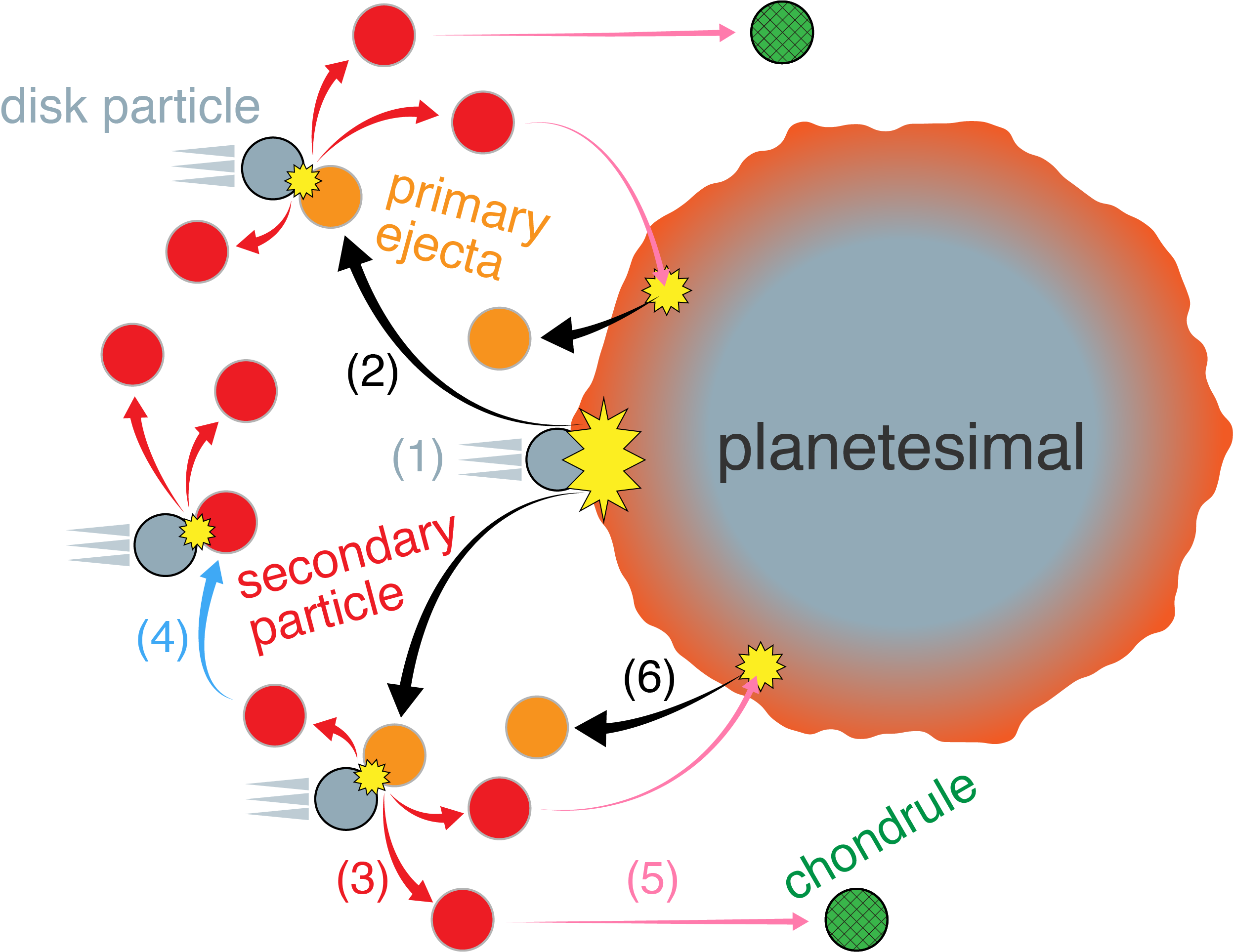}
\caption{
Schematic illustration of the heavy bombardment of a planetesimal by debris dust and the formation of a dense dust cloud.
(1) Within the debris-dust layer, a fraction of the debris particles directly collide with the planetesimal surface.
(2) These impacts generate primary ejecta around the planetesimal.
(3) Another fraction of the incoming debris collides with the primary ejecta before reaching the surface, producing secondary particles that form a dense dust cloud.
(4) Some of the incoming debris also collides with the secondary particles within this dense cloud, generating additional secondary particles.
(5) Some of the dust-cloud particles eventually fall back onto the planetesimal surface, whereas a fraction escapes laterally from the vicinity of the planetesimal and ultimately forms chondrules.
(6) Impacts of fallback cloud particles onto the planetesimal surface also generate primary ejecta.
}
\label{fig:3}
\end{figure}

As discussed above, a situation in which high-speed eccentric planetesimals penetrate a geometrically thin debris disk is expected to arise after gas in the inner region of the solar protoplanetary disk has dispersed.
Here, we schematically illustrate the heavy bombardment of planetesimals by debris dust, which leads to the formation of a dense dust cloud around the planetesimals and to chondrule production through multiple collisions (see Figure~\ref{fig:3}).

Once an eccentric planetesimal enters the debris-dust layer, a fraction of the debris particles directly collide with the planetesimal surface.
These impacts generate primary ejecta around the planetesimal, while another fraction of the incoming debris collides with the primary ejecta before reaching the surface.
Collisions between the primary ejecta and the incoming debris generate secondary particles, which form a dense dust cloud surrounding the planetesimal.
Some of the incoming debris also collides with the secondary particles within the cloud, producing additional secondary particles.
These collisional processes are highly energetic, and the particles in the dust cloud are expected to be molten.
Some of the dust-cloud particles eventually fall back onto the planetesimal surface, forming a melt ocean, whereas a fraction escapes laterally from the vicinity of the planetesimal.
These escaping particles gradually cool and solidify after passing the planetesimal, ultimately becoming chondrules.
The fallback of cloud particles onto the planetesimal surface regenerates primary ejecta, which are subsequently converted into cloud particles.

In our scenario, the debris initially present in the disk is assumed to originate from two sources: (i) fragments produced by planetesimal collisions and (ii) dust aggregates that grew in the protosolar disk.
Refractory condensates formed in the protosolar disk, such as CAIs, would also have been present as part of this primordial debris.
As chondrule formation proceeds, a substantial fraction of the dust component in the debris disk would be progressively replaced by (iii) chondrules leaking from dust clouds.
The subsequent gravitational instability of the chondrule-hosting debris disk would then lead to the formation of chondritic planetesimals.

This scenario may qualitatively explain the existence of chondrules that experienced multiple heating events, as recorded by relict grains and chondrule fragments enclosed in chondrules \citep[e.g.,][]{2024SSRv..220...69M}.
Particles leaking from the dust cloud become chondrules and are supplied to the debris disk.
Some of these chondrules floating within the debris disk may then be reinjected into dust clouds surrounding eccentric planetesimals.
During this process, some of the injected chondrules do not reach the molten planetesimal surface, but instead collide with primary ejecta or cloud particles and become incorporated into the dust cloud.
Some of these chondrules may avoid complete melting, allowing relict grains or chondrule fragments to survive.
Such relict grains and chondrule fragments could remain in the dust cloud and later merge with other chondrules, thereby forming chondrules that preserve evidence of multiple heating events \citep[e.g.,][]{2008GeCoA..72.5530R}.

Another possible line of evidence for chondrule recycling is the size dependence of oxygen isotopic compositions among chondrules in ordinary chondrites \citep[e.g.,][]{2024GeCoA.371...52M}.
Specifically, small chondrules with diameters of $\lesssim 0.3~{\rm mm}$ tend to be more $^{16}$O-rich than larger chondrules.
Although the origin of this size dependence remains debated \citep[e.g.,][]{hanai2025oxygen}, one possible explanation is that these small chondrules formed earlier and subsequently served as precursors to larger chondrules through reprocessing in a relatively $^{16}$O-poor environment \citep[e.g.,][]{2024GeCoA.371...52M}.
If the initial dust components and planetesimals in the debris disk had distinct oxygen isotopic compositions, the temporal evolution of the oxygen isotopic composition of the chondrule-forming environment could be explained within the framework of the chondrule recycling hypothesis.
This possibility should be investigated quantitatively in future studies.

\section{Chondrule Production Rate}
\label{sec:rate}

\begin{figure}[]
\centering
\includegraphics[width = \columnwidth]{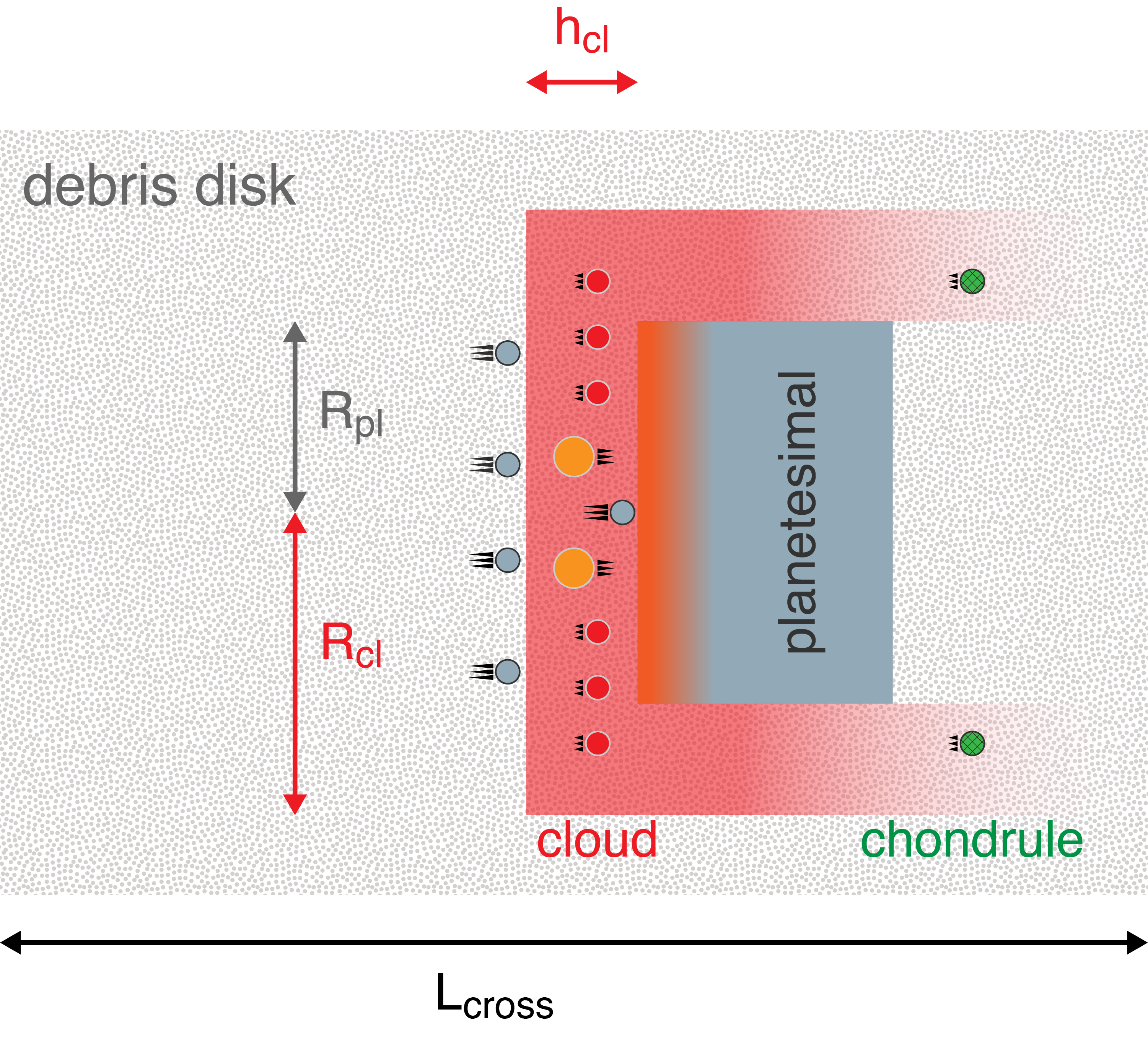}
\caption{
Schematic of a planetesimal and its surrounding dust cloud embedded in the debris disk.
For simplicity, the planetesimal is assumed to be cylindrical, and the structure of the dust cloud is shown in a side view.
We consider three species of dust particles: incoming debris-disk particles ({\bf in}, gray circles), primary ejecta ({\bf ej}, large orange circles), and secondary cloud particles ({\bf cl}, red circles).
A fraction of the secondary cloud particles escapes laterally from the vicinity of the planetesimal and eventually becomes chondrules (cross-hatched green circles).
}
\label{fig:4}
\end{figure}

To assess the viability of the proposed chondrule formation mechanism, namely heavy bombardment of planetesimals by debris dust, we perform a conceptual simulation of the temporal evolution of the dense dust cloud surrounding a planetesimal.
We present a schematic illustration of a planetesimal and its surrounding dust cloud embedded in the debris disk in Figure~\ref{fig:4}.
For simplicity, the planetesimal is assumed to be cylindrical, and the structure of the dust cloud is shown in side view.
The thickness of the dust cloud is denoted by $h_{\rm cl}$.
We assume that the radius of the dust cloud, $R_{\rm cl}$, is given by $R_{\rm cl} = R_{\rm pl} + h_{\rm cl}$.
The path length of a planetesimal through the debris disk during a single crossing is denoted by $L_{\rm cross}$, and its residence time within the disk is $t_{\rm cross} = L_{\rm cross} / v_{\rm pl}$.
In this study, we consider only the rocky component and neglect the presence of ice (see Appendix~\ref{app:ice}).

We consider three species of dust particles: incoming debris-disk particles ({\bf in}), primary ejecta ({\bf ej}), and secondary cloud particles ({\bf cl}).
The surface density of the primary ejecta, integrated along the direction of the planetesimal's motion, is denoted by $\Sigma_{\rm ej}$, and that of the secondary cloud particles by $\Sigma_{\rm cl}$.
Similarly, the momenta of the primary ejecta and the secondary cloud particles per unit area are denoted by $p_{\rm ej}$ and $p_{\rm cl}$, respectively.
The mean velocity and density of the secondary cloud particles are given by $v_{\rm cl} = p_{\rm cl} / \Sigma_{\rm cl}$ and $\rho_{\rm cl} = \Sigma_{\rm cl} / h_{\rm cl}$, respectively.
Similarly, the density of the primary ejecta is given by $\rho_{\rm ej} = \Sigma_{\rm ej} / h_{\rm cl}$.
The cloud thickness $h_{\rm cl}$ is determined by the collisional deceleration of the primary ejecta (see Appendix~\ref{app:eq}).

The temporal evolution of $\Sigma_{\rm cl}$ and $\Sigma_{\rm ej}$ is given by the following equations:
\begin{eqnarray}
\dot{\Sigma}_{\rm cl} & = & \dot{\Sigma}_{{\rm in} \to {\rm cl}} + \dot{\Sigma}_{{\rm ej} \to {\rm cl}} - \dot{\Sigma}_{{\rm cl} \to {\rm ex}}, \\
\dot{\Sigma}_{\rm ej} & = & \dot{\Sigma}_{{\rm in} \to {\rm ej}} + \dot{\Sigma}_{{\rm cl} \to {\rm ej}} - \dot{\Sigma}_{{\rm ej} \to {\rm cl}}.
\end{eqnarray}
Here, $\dot{\Sigma}_{{\rm in} \to {\rm cl}}$ denotes the mass conversion rate from the {\bf in} to the {\bf cl} component, $\dot{\Sigma}_{{\rm ej} \to {\rm cl}}$ denotes that from the {\bf ej} to the {\bf cl} component, and $\dot{\Sigma}_{{\rm cl} \to {\rm ex}}$ represents the removal rate of the {\bf cl} component from the dust cloud system.
Primary ejecta are generated when incoming disk particles or fallback cloud particles impact the planetesimal surface.
The corresponding mass conversion rates, $\dot{\Sigma}_{{\rm in} \to {\rm ej}}$ and $\dot{\Sigma}_{{\rm cl} \to {\rm ej}}$, are calculated using the crater-ejecta scaling law \citep[e.g.,][]{2011Icar..211..856H}.
The temporal evolution of $p_{\rm ej}$ and $p_{\rm cl}$ is also given by similar equations:
\begin{eqnarray}
\dot{p}_{\rm cl} & = & \dot{p}_{{\rm in} \to {\rm cl}} + \dot{p}_{{\rm ej} \to {\rm cl}} - \dot{p}_{{\rm cl} \to {\rm ex}}, \\
\dot{p}_{\rm ej} & = & \dot{p}_{{\rm in} \to {\rm ej}} + \dot{p}_{{\rm cl} \to {\rm ej}} - \dot{p}_{{\rm ej} \to {\rm cl}}.
\end{eqnarray}
The specific forms of the terms on the right-hand sides of these evolution equations are given in Appendix~\ref{app:eq}.
We also evaluate the chondrule production rate, $\dot{M}_{\rm ch}$, using
\begin{equation}
\dot{M}_{\rm ch} = {\left( S_{\rm cl} - S_{\rm pl} \right)} \dot{\Sigma}_{{\rm cl} \to {\rm ex}},
\end{equation}
where $S_{\rm cl} = \pi {R_{\rm cl}}^{2}$ and $S_{\rm pl} = \pi {R_{\rm pl}}^{2}$ are the cross sections of the dust cloud and the planetesimal, respectively.

\begin{figure*}[]
\centering
\includegraphics[width = \textwidth]{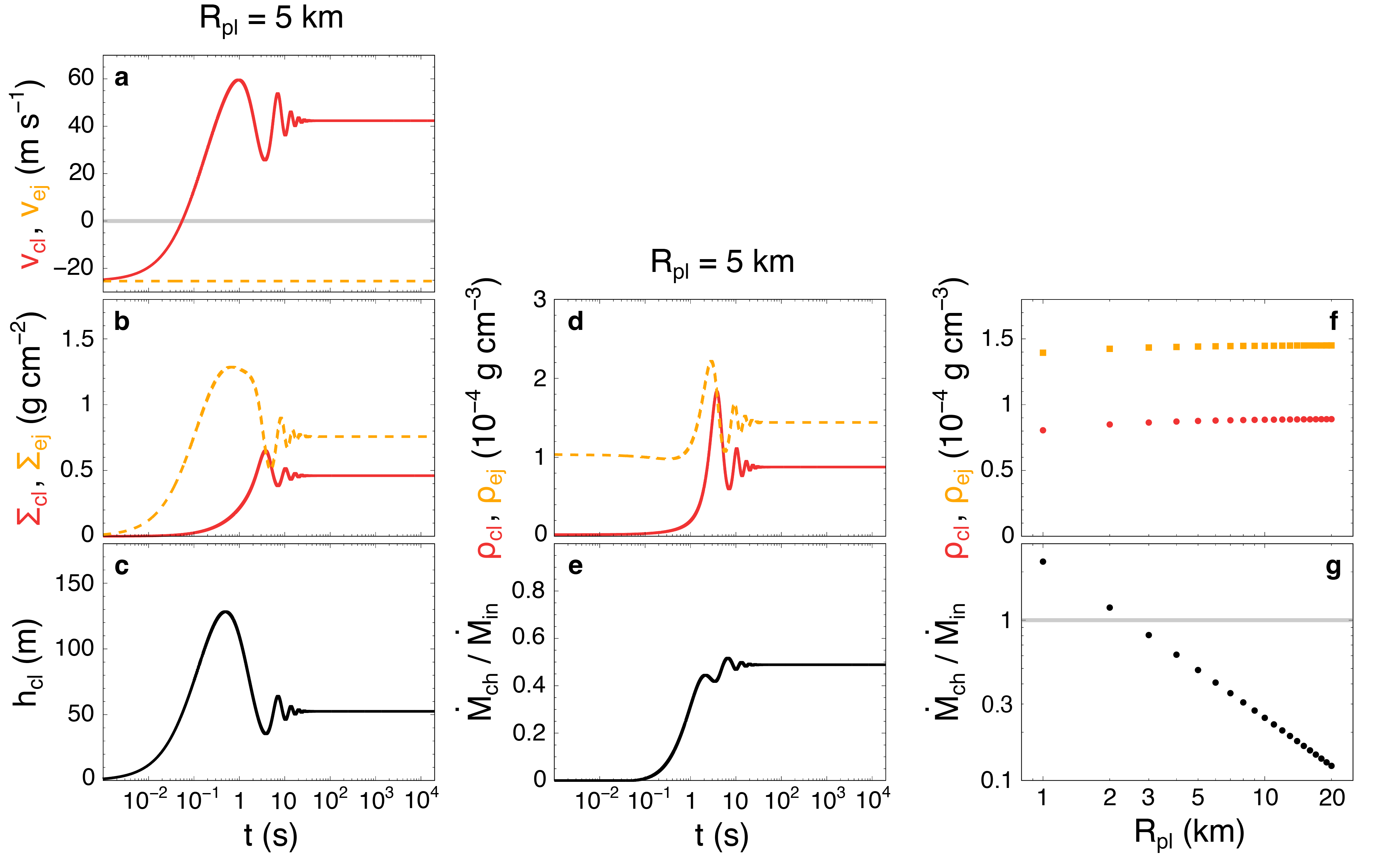}
\caption{
{\bf a}--{\bf e}: Temporal evolution of the dust-cloud system for $R_{\rm pl} = 5~{\rm km}$.
{\bf a}: $v_{\rm cl}$ (red solid line) and $v_{\rm ej}$ (orange dashed line).
{\bf b}: $\Sigma_{\rm cl}$ (red solid line) and $\Sigma_{\rm ej}$ (orange dashed line).
{\bf c}: $h_{\rm cl}$.
{\bf d}: $\rho_{\rm cl}$ (red solid line) and $\rho_{\rm ej}$ (orange dashed line).
{\bf e}: $\dot{M}_{\rm ch} / \dot{M}_{\rm in}$.
{\bf f} and {\bf g}: Dependence on $R_{\rm pl}$.
{\bf f}: Steady-state values of $\rho_{\rm cl}$ (red circles) and $\rho_{\rm ej}$ (orange squares).
{\bf g}: Steady-state values of $\dot{M}_{\rm ch} / \dot{M}_{\rm in}$.
}
\label{fig:6}
\end{figure*}

Figures~\ref{fig:6}{\bf a}--\ref{fig:6}{\bf c} show the temporal evolution of the velocities ($v_{\rm cl}$ and $v_{\rm ej}$), surface densities ($\Sigma_{\rm cl}$ and $\Sigma_{\rm ej}$), and cloud thickness ($h_{\rm cl}$) during a single crossing of the debris disk by a planetesimal with $R_{\rm pl} = 5~{\rm km}$.
Here, we set $L_{\rm cross} = 1 \times 10^{5}~{\rm km}$, $v_{\rm pl} = 5~{\rm km}~{\rm s}^{-1}$, and $\rho_{\rm d} = 1 \times 10^{-8}~{\rm g}~{\rm cm}^{-3}$.
The residence time of the planetesimal in the debris disk is $t_{\rm cross} = L_{\rm cross} / v_{\rm pl} = 2 \times 10^{4}~{\rm s}$, and the dust cloud surrounding the planetesimal disperses immediately after the planetesimal passes through the debris disk.
We find that the dust cloud reaches its equilibrium structure within several tens of seconds after the planetesimal enters the debris disk.

The temporal evolution of the densities $\rho_{\rm cl}$ and $\rho_{\rm ej}$ is shown in Figure~\ref{fig:6}{\bf d}.
For this set of parameters, the equilibrium densities are $\rho_{\rm cl} \sim \rho_{\rm ej} \sim 10^{-4}~{\rm g}~{\rm cm}^{-3}$.
This value satisfies cosmochemical constraints on the chondrule-forming environment, particularly those based on volatile-element retention within chondrules \citep[e.g.,][]{2008Sci...320.1617A, 2013GeCoA.112..226F}.

The mass influx of debris dust into the dust-cloud system is given by $\dot{M}_{\rm in} = S_{\rm cl} \rho_{\rm d} v_{\rm pl}$, and the chondrule production rate is denoted by $\dot{M}_{\rm ch}$.
Here we define the conversion efficiency as $\dot{M}_{\rm ch} / \dot{M}_{\rm in}$, which is shown in Figure~\ref{fig:6}{\bf e}.
We find that the conversion efficiency reaches $\approx 50\%$ in the steady state at $t \gg 10~{\rm s}$.

The steady-state structure of the dust-cloud system depends on parameters such as $R_{\rm pl}$.
Figure~\ref{fig:6}{\bf f} shows the equilibrium values of $\rho_{\rm cl}$ and $\rho_{\rm ej}$ as a function of $R_{\rm pl}$.
For planetesimals with $1~{\rm km} \le R_{\rm pl} \le 20~{\rm km}$, both $\rho_{\rm cl}$ and $\rho_{\rm ej}$ are approximately $10^{-4}~{\rm g}~{\rm cm}^{-3}$ and are nearly independent of $R_{\rm pl}$.
This suggests that, in the debris-dust bombardment scenario, chondrule formation in a highly dust-enriched environment can be achieved over a wide range of $R_{\rm pl}$.
The dependence of the steady-state structure on other key parameters, such as the ejecta radius and the minimum ejecta velocity, is discussed in Appendices \ref{app:rad} and \ref{app:v_ej}.

Finally, we show the steady-state conversion efficiency as a function of $R_{\rm pl}$ (Figure~\ref{fig:6}{\bf g}).
For planetesimals with $3~{\rm km} \le R_{\rm pl} \le 20~{\rm km}$, $\dot{M}_{\rm ch} / \dot{M}_{\rm in}$ is approximately inversely proportional to $R_{\rm pl}$ and ranges from 10\% to 100\%.
This estimate is consistent with our hypothesis that a fraction of the incoming debris dust is deposited onto the planetesimal surface and forms a melt ocean.
We also note that the conversion efficiency can exceed 100\%, depending on the choice of $R_{\rm pl}$.
Indeed, in the present case, we find that $\dot{M}_{\rm ch} / \dot{M}_{\rm in} > 1$ for $R_{\rm pl} \le 2~{\rm km}$.
When $\dot{M}_{\rm ch} / \dot{M}_{\rm in} > 1$, planetesimals may undergo net collisional erosion, thereby supplying mass to the debris disk.


If planetesimals had differentiated into silicate and metal components, their collisional erosion could potentially produce dust clouds that are extremely depleted or enriched in metal.
Such dust clouds would likely be inconsistent with the typical chemical compositions of chondrules in the Solar System, except possibly for non-porphyritic chondrules in CH and CB chondrites \citep[e.g.,][]{2002M&PS...37.1451K}.
However, simple thermal evolution calculations show that the peak temperature of planetesimals with $R_{\rm pl} \lesssim 2~{\rm km}$ does not reach $1500~{\rm K}$, irrespective of their accretion age (see Appendix~\ref{app:T_peak}).
Therefore, collisional erosion of differentiated planetesimals is unlikely to play a major role in our chondrule formation model.


\section{Heat balance analysis}
\label{sec:heat}

\begin{figure}[]
\centering
\includegraphics[width = \columnwidth]{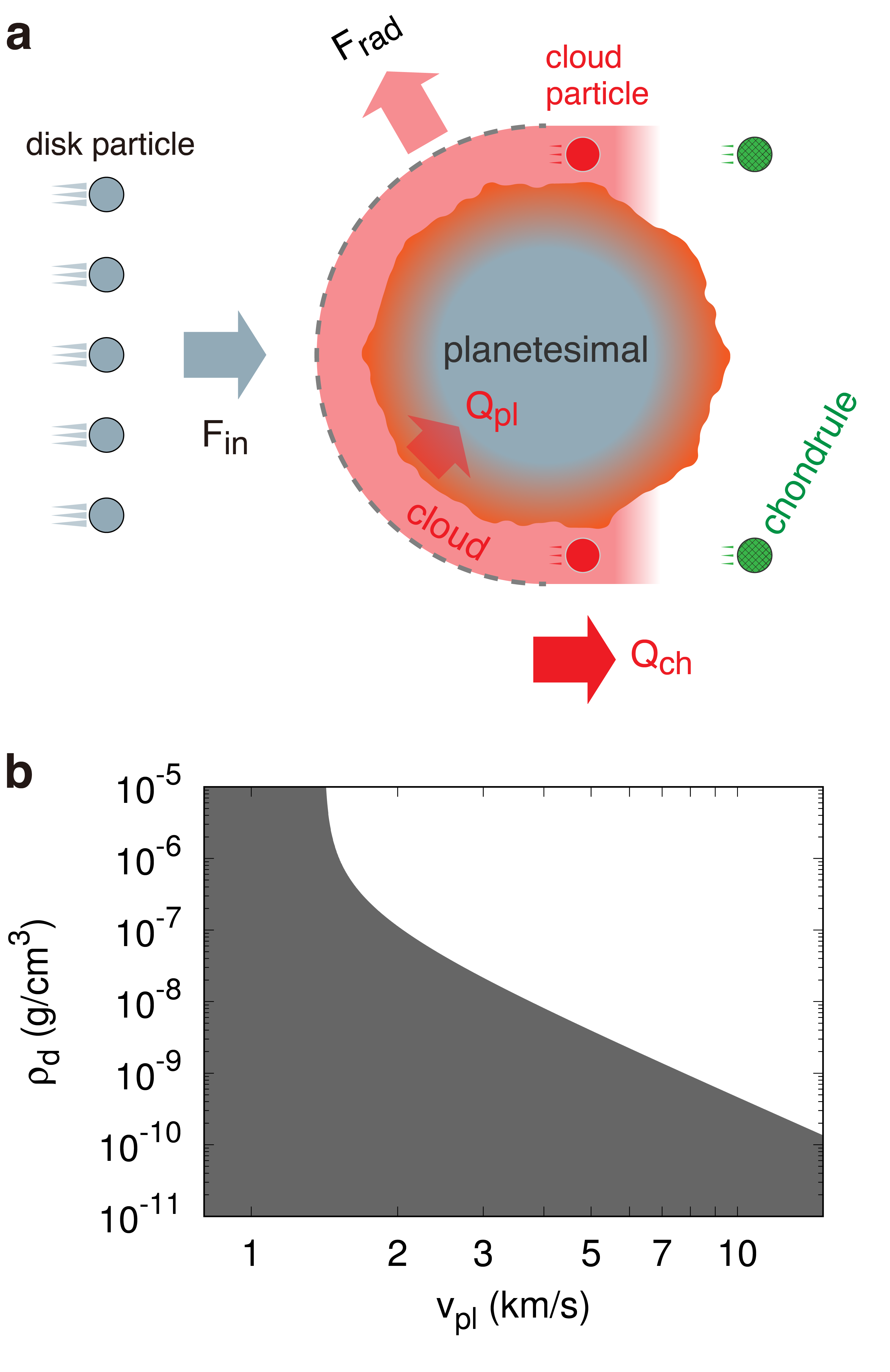}
\caption{
{\bf a}: Schematic illustration of the heat balance around the planetesimal.
The kinetic energy influx is denoted by $F_{\rm in}$, and the radiative energy efflux by $F_{\rm rad}$.
The thermal energy consumption rate associated with chondrule formation is denoted by $Q_{\rm ch}$, and that at the planetesimal surface by $Q_{\rm pl}$.
{\bf b}: Parameter space in the $(v_{\rm pl}, \rho_{\rm d})$ plane that satisfies the conditions for maintaining a melt ocean on the planetesimal surface and producing a sufficient amount of chondrules (white region).
}
\label{fig:5}
\end{figure}

We also evaluate whether a melt ocean can be maintained during a debris-disk crossing using a heat-balance analysis.
Figure~\ref{fig:5}{\bf a} presents a schematic illustration of our heat-balance analysis around the planetesimal.
Assuming that the photosphere of a planetesimal surrounded by a dust cloud is located at a height $h_{\rm cl}$ above the surface (i.e., at a distance $r_{\rm cl}$ from the planetesimal's center), the radiative energy efflux, $F_{\rm rad}$, is given by
\begin{equation}
F_{\rm rad} \sim 4 S_{\rm cl} \sigma_{\rm SB} {T_{\rm cl}}^{4},
\end{equation}
where $T_{\rm cl}$ denotes the temperature at the cloud photosphere.
For simplicity, we assume that the dust cloud envelops the entire planetesimal.
The high-speed incoming debris dust delivers kinetic energy to the dust cloud, and the corresponding energy influx, $F_{\rm in}$, is
\begin{equation}
F_{\rm in} \approx \frac{1}{2} S_{\rm cl} \rho_{\rm d} {v_{\rm pl}}^{3}.
\end{equation}
Cloud particles that leak out and subsequently become chondrules remove thermal energy from the dust cloud.
The corresponding energy consumption rate, $Q_{\rm ch}$, is given by
\begin{equation}
Q_{\rm ch} \approx \dot{M}_{\rm ch} c_{\rm heat} {\Delta T},
\end{equation}
where $c_{\rm heat} = 1000~{\rm J}~{\rm kg}^{-1}~{\rm K}^{-1}$ is the specific heat capacity of chondrules, and ${\Delta T}$ is the temperature difference between the escaping chondrules and the incoming debris dust.
Assuming a background temperature of the debris disk of $\lesssim 500~{\rm K}$ and a dust-cloud temperature suitable for chondrule formation of $\sim 1500~{\rm K}$, we estimate ${\Delta T} \sim 1000~{\rm K}$.
In addition, thermal energy may be consumed at the planetesimal surface to sustain the melt ocean; we denote the corresponding energy consumption rate by $Q_{\rm pl}$.

We evaluate the optical depth of the dust cloud, $\tau_{\rm cl}$.
Here, we neglect the contribution of the primary ejecta to the optical depth of the dust-cloud system, because the primary ejecta are assumed to have larger grain sizes than the cloud particles and thus lower opacity.
As the cloud particles are submillimeter-sized grains, their opacity is assumed to follow the geometrical optics approximation.
The cross section of the cloud particles per unit mass, $\kappa_{\rm cl}$, is given by $\kappa_{\rm cl} \approx A_{\rm cl} / m_{\rm cl} \approx 25~{\rm cm}^{2}~{\rm g}^{-1}$.
The optical depth is then given by $\tau_{\rm cl} \approx \kappa_{\rm cl} \Sigma_{\rm cl} \approx 10{\left( {\Sigma_{\rm cl}} / {0.4~{\rm g}~{\rm cm}^{-2}} \right)}$.
Our estimate of $\tau_{\rm cl}$ indicates that the dust cloud surrounding the planetesimal is indeed optically thick under our fiducial conditions (see Figure~\ref{fig:6}{\bf b}).
The temperature at the base of the dust cloud is expected to be comparable to that of the planetesimal surface.
If the planetesimal sustains a (partially molten) melt ocean, the corresponding temperature would be approximately $1500~{\rm K}$.
The temperature at the photosphere of the dust cloud, $T_{\rm cl}$, would be lower than that at the planetesimal surface, although its actual value depends on the details of heat transfer and kinetic processes within the dust cloud.
For the following discussion, we adopt $T_{\rm cl} \approx 1000~{\rm K}$.

When the dust cloud is in a steady state, the heat-balance equation, $F_{\rm in} = F_{\rm rad} + Q_{\rm ch} + Q_{\rm pl}$, must be satisfied.
To maintain a melt ocean on the planetesimal surface, $Q_{\rm pl} > 0$ is required.
In addition, to produce a sufficient amount of chondrules, $\dot{M}_{\rm ch}$ should be comparable to $\dot{M}_{\rm in} = S_{\rm cl} \rho_{\rm d} v_{\rm pl}$.
Under these conditions, we derive the requirement for $Q_{\rm pl} > 0$ as follows:
\begin{equation}
\frac{1}{2} \rho_{\rm d} {v_{\rm pl}}^{3} \gtrsim 4 \sigma_{\rm SB} {T_{\rm cl}}^{4} + \rho_{\rm d} v_{\rm pl} c_{\rm heat} {\Delta T}.
\label{eq:balance}
\end{equation}
The white region in Figure~\ref{fig:5}{\bf b} shows the parameter space that satisfies Inequality~\eqref{eq:balance}.
We note that this heat-balance analysis is independent of the choice of $R_{\rm pl}$.
For $v_{\rm pl} \gtrsim 2~{\rm km}~{\rm s}^{-1}$, Inequality~\eqref{eq:balance} can be rewritten as
\begin{equation}
\rho_{\rm d} \gtrsim 4 \times 10^{-9}~{\left( \frac{v_{\rm pl}}{5~{\rm km}~{\rm s}^{-1}} \right)}^{-3}~{\rm g}~{\rm cm}^{-3}.
\end{equation}
The required condition on $(v_{\rm pl}, \rho_{\rm d})$ can indeed be satisfied in a Solar System debris disk hosting eccentric planetesimals, as estimated from Equation~\eqref{eq:rho_d}.


When considering the chondrule-forming environment, it is necessary to take into account not only temperature but also pressure.
Chondrules are thought to have formed in environments with substantially elevated partial pressures of silicate vapor \citep[e.g.,][]{2000GeCoA..64..339E, 2006E&PSL.251..232L, 2008Sci...320.1617A}.
Indeed, to melt silicate materials in the disk and form chondrules while avoiding substantial sodium loss, it is desirable to have (i) silicate dust enrichment by a factor of $\gtrsim 10^{2}$ relative to solar composition and (ii) high pressures of order $\sim 10^{2}~{\rm Pa}$ \citep[e.g.,][]{2000GeCoA..64..339E}.

In our model, chondrule formation occurs in an H$_{2}$-gas-depleted, dust-rich environment, and therefore condition (i) is clearly satisfied.
Here we further show that condition (ii) can also be satisfied.
The temperature of the dust cloud is high enough to induce at least partial melting of silicate materials, and therefore silicate vapor is expected to be present within the cloud.
The dust cloud is subject to ram pressure due to the influx of debris-disk particles, and we assume that the silicate vapor pressure is comparable to the ram pressure acting on the dust cloud.
The ram pressure acting on the dust cloud, $P_{\rm cl}$, can be estimated as
\begin{eqnarray}
P_{\rm cl} & \approx & \frac{1}{2} \rho_{\rm d} {v_{\rm pl}}^{2} \nonumber \\
           & \sim    & 100 {\left( \frac{\rho_{\rm d}}{1 \times 10^{-8}~{\rm g}~{\rm cm}^{-3}} \right)} {\left( \frac{v_{\rm pl}}{5~{\rm km}~{\rm s}^{-1}} \right)}^{2}~{\rm Pa}.
\end{eqnarray}
Thus, the dust cloud in our model can realize an environment with high silicate vapor pressure ($\sim 10^{2}~{\rm Pa}$), favorable for chondrule formation.


\section{Chronology of Chondrule Formation}
\label{sec:Jupiter}

As discussed above, our proposed scenario---the high-velocity entry of eccentric planetesimals into a debris-dust disk---may provide a mechanism for producing chondrule-like igneous objects.
In the early Solar System, the formation of giant planets such as Jupiter would have increased the orbital eccentricities of planetesimals.
Recent studies of the dichotomy in nucleosynthetic isotopic anomalies between inner- and outer-Solar-System materials have suggested that proto-Jupiter may have formed within $\lesssim 1~{\rm Myr}$ after CAI formation \citep[e.g.,][]{2017PNAS..114.6712K}, potentially altering the disk structure by opening a gap and thereby inhibiting the radial migration of dust particles.
Planet-formation models that support such rapid proto-Jupiter formation have also been proposed.
For example, \citet{2021ApJ...922...16K} showed that dust grains in the solar protoplanetary disk can grow to form the proto-Jupiter core on a timescale of several $10^{5}$ years.
Proto-Jupiter could then grow to its present mass on a timescale of $10^{5}$--$10^{6}$ years through the accretion of disk gas \citep[e.g.,][]{2016ApJ...823...48T}.

Such rapid formation of Jupiter may also be consistent with the inferred onset of chondrule formation based on recent high-precision $^{26}{\rm Al}$--$^{26}{\rm Mg}$ dating of chondrules.
We note, however, that the onset of chondrule formation remains under debate, as it depends on the assumption of the spatial homogeneity of $^{26}{\rm Al}$.
If the short-lived radionuclide $^{26}{\rm Al}$ was spatially homogeneous throughout the solar protoplanetary disk, with an initial abundance of $^{26}{\rm Al} / ^{27}{\rm Al} \approx 5 \times 10^{-5}$ \citep[e.g.,][]{2023Icar..40215607D}, chondrule formation in ordinary chondrites would have begun at $\approx 1.8~{\rm Myr}$ after CAI formation \citep[e.g.,][]{2021GeCoA.293..103S}.
In contrast, if $^{26}{\rm Al}$ was heterogeneously distributed \citep[e.g.,][]{2025ApJ...979L..29I}, the inferred onset age shifts to $\approx 0.4~{\rm Myr}$ after CAI formation.
These inferred onset ages are also consistent with the early formation of the proto-Jupiter core \citep[$\lesssim 1~{\rm Myr}$ after CAI formation; e.g.,][]{2017PNAS..114.6712K}.

The duration of chondrule formation has also been investigated using $^{26}{\rm Al}$--$^{26}{\rm Mg}$ dating.
Recent high-precision $^{26}{\rm Al}$--$^{26}{\rm Mg}$ dating has enabled detailed discussion of the duration of chondrule formation, showing that chondrules in many ordinary chondrites formed within a narrow time window of approximately $0.4~{\rm Myr}$ \citep[e.g.,][]{2021GeCoA.293..103S, 2022GeCoA.324..312S}.
In our scenario, the timescale over which the debris-dust layer is heated by eccentric planetesimals is $\tau_{\rm pl} \sim 10^{5}~{\rm yr}$ (Equation~\eqref{eq:tau_pl}), which can account for chondrule formation within such a short interval.

\section{Summary}

In this study, we proposed a new chondrule formation scenario in which eccentric planetesimals are heavily bombarded by debris dust in the gas-depleted inner protosolar disk (see Section~\ref{sec:overview}).
After the formation of proto-Jupiter, gas accretion onto proto-Jupiter and disk winds would have depleted the inner disk gas, leaving behind a geometrically thin, dust-rich debris layer.
Planetesimals excited onto eccentric orbits repeatedly penetrate this dust-rich disk at velocities of several km/s, producing molten ejecta and dense dust clouds around the planetesimals through successive collisions with incoming debris-disk particles.
A fraction of the molten cloud particles escapes from the vicinity of the planetesimals and solidifies into chondrules.

Our dust-cloud model indicates that the cloud density could reach $\sim 10^{-4}~{\rm g}~{\rm cm}^{-3}$ and that the silicate vapor pressure in the cloud could reach $\sim 10^{2}~{\rm Pa}$ owing to the ram pressure exerted by incoming debris dust (see Sections~\ref{sec:rate} and \ref{sec:heat}, respectively).
Our model thus provides suitable conditions for chondrule formation \citep[e.g.,][]{2000GeCoA..64..339E, 2006E&PSL.251..232L, 2008Sci...320.1617A}.
The conversion efficiency from incoming debris dust to escaping chondrules ranges from about 10\% to 100\% for planetesimals with radii of a few to tens of kilometers (Section~\ref{sec:rate}).
The characteristic processing timescale of the debris layer is about $10^{5}$~years, consistent with the short duration of chondrule formation inferred for ordinary chondrites \citep[e.g.,][]{2021GeCoA.293..103S, 2022GeCoA.324..312S} (see Section~\ref{sec:Jupiter}).
Furthermore, our model can explain the presence of chondrules that experienced multiple heating events prior to accretion into their parent bodies \citep[e.g.,][]{2007E&PSL.257..274R, 2018crpd.book..192T, 2021ApJ...910...70V, 2024SSRv..220...69M}.

Our scenario therefore links Jupiter formation, inner-disk gas depletion, planetesimal excitation, and chondrule formation within a single framework.
In this picture, chondrule formation naturally arises as a consequence of the evolution of the inner protosolar disk.
The present study uses several order-of-magnitude estimates to evaluate the viability of this scenario, which should be tested with more detailed numerical simulations in future studies.
Moreover, the outcomes of high-velocity collisions involving liquid silicates remain poorly understood, despite the key role of these collisions in our scenario.
We therefore plan to investigate the underlying physical processes through laboratory experiments and to refine our dust-cloud model based on the experimental results.

\begin{acknowledgments}

The authors thank Shota Sato and Dr.~Alexander N.~Krot for carefully reading the original manuscript and for providing helpful comments.
SA was supported by JSPS KAKENHI Grants JP24K17118, JP24KK0072, JP25K00025, and JP25H00678; TK by JP23K03483; TU by JP22K18741, JP24H00259, and JP25K07382; MN by JP21K03650; and HK by JP24K00690 and JP25K01055.

\end{acknowledgments}

\begin{contribution}

HT and SA developed the idea for the study.
SA performed the numerical simulations and prepared the original draft.
All authors contributed to the interpretation of the results and to writing the paper.


\end{contribution}

%



\appendix

\section{Random velocity of the planetesimals}
\label{app:v_pl}

We set $v_{\rm pl} = 5~{\rm km}~{\rm s}^{-1}$ as the fiducial value, which is broadly consistent with the typical random velocity in the present-day asteroid belt \citep[e.g.,][]{1994Icar..107..255B}.
\citet{2014ApJ...794L...7N, 2019ApJ...871..110N} showed that, in the minimum-mass solar nebula \citep{1981PThPS..70...35H}, a planetesimal with $R_{\rm pl} \sim 10~{\rm km}$ initially located at $r \sim 4~{\rm au}$ can reach $v_{\rm pl} \sim 5~{\rm km}~{\rm s}^{-1}$ under Jupiter's gravitational perturbation.
In our gas-depleted debris-disk model, weaker gas drag may allow even smaller planetesimals to attain such large $v_{\rm pl}$ values.
However, because the surface density of the debris disk is lower than that of the minimum-mass solar nebula, eccentricity excitation by secular resonances may also be less effective \citep[e.g.,][]{2019ApJ...871..110N}.
Moreover, neither Jupiter's mass nor the disk mass remains constant over time: Jupiter grows by accreting disk gas, while the disk mass decreases over time.
Future simulations of planetesimal orbital evolution that incorporate the evolution of both Jupiter's mass and the disk surface density are essential for assessing the plausible $v_{\rm pl}$ achievable in chondrule-forming environments.

\section{Chondrule formation beyond the snow line}
\label{app:ice}

In this study, for simplicity, we consider only the rocky component and neglect the presence of ice.
Therefore, although our current model may explain the origin of chondrules in ordinary and enstatite chondrites, it is presently difficult to account for the origin of chondrules in carbonaceous chondrites.
Chondrules in carbonaceous chondrites are known to exhibit a correlation between redox state and oxygen isotopic composition, which is generally interpreted as reflecting the influence of ice in the chondrule-forming region \citep[e.g.,][]{2018crpd.book..192T}.
Moreover, the fact that many carbonaceous chondrites have experienced aqueous alteration \citep[e.g.,][]{2025SSRv..221...11L} provides further evidence that ice was present in the chondrule-forming environment of carbonaceous chondrites.
If ice was present in the incoming debris dust, water vapor produced by its sublimation should also have been present in the dust cloud.
Future work should incorporate the effects of water vapor and examine whether the origin of chondrules in carbonaceous chondrites can also be explained within our heavy-bombardment scenario.

Assuming that $^{26}{\rm Al}$ was spatially homogeneous throughout the solar protoplanetary disk and that its initial abundance was $^{26}{\rm Al}/^{27}{\rm Al} \approx 5 \times 10^{-5}$ \citep[e.g.,][]{2023Icar..40215607D}, chondrule formation in carbonaceous chondrites other than Renazzo-type (CR) chondrites would have begun at $\approx 2.2~{\rm Myr}$ after CAI formation and lasted for approximately $0.6~{\rm Myr}$ \citep{2022GeCoA.322..194F}.
This suggests that chondrules in carbonaceous chondrites, which formed beyond the snow line, formed later than those in ordinary chondrites, which formed inside the snow line.
By contrast, if gravitational perturbations from Jupiter excited the eccentricities of planetesimals interior to Jupiter's orbit, the heating region associated with eccentric planetesimals may have migrated inward over time.
This trend, however, is opposite to the inferred age trend of chondrule formation.
This suggests that chondrules in carbonaceous chondrites may have formed exterior to Jupiter's orbit.

Saturn likely formed after Jupiter by accreting gas from the solar protoplanetary disk, and its growth probably contributed to gas depletion beyond Jupiter's orbit.
Following Saturn's formation, gravitational perturbations from Jupiter and Saturn could have excited the eccentricities of planetesimals located between their orbits.
Under such conditions, heavy bombardment of planetesimals by debris dust may have occurred beyond Jupiter's orbit.
Such a scenario may help explain the age difference between chondrules in carbonaceous and ordinary chondrites.
Recent studies of the nucleosynthetic isotope dichotomy between carbonaceous and non-carbonaceous chondrites have discussed similar scenarios in which carbonaceous chondrites and their chondrules formed exterior to Jupiter's orbit \citep[e.g.,][]{2020SSRv..216...55K, 2017PNAS..114.6712K}.
Further discussion of the formation environment and mechanism of chondrules in carbonaceous chondrites is therefore essential.

\section{Governing equations of the dust-cloud system}
\label{app:eq}

We assume that the radius of the primary ejecta, $R_{\rm ej}$, is sufficiently larger than that of the incoming debris dust, for which we adopt $R_{\rm d} = 0.1~{\rm mm}$.
For the fiducial case shown in Figure~\ref{fig:6}, we set $R_{\rm ej} = 5~{\rm mm}$.
We also examine the effect of $R_{\rm ej}$ on the dust-cloud structure in Figure~\ref{fig:e1}.
Additionally, we assume that the radius of the cloud particles is equal to $R_{\rm d}$, because cloud particles leaking from the dust cloud eventually become debris dust.

In reality, cloud particles are likely to experience multiple collisions within the dust cloud, and their size distribution would be determined by the balance between collisional fragmentation and coalescence.
The existence of compound chondrules in chondrites is also evidence that multiple collisions occurred within the dust cloud \citep[e.g.,][]{2016Icar..276..102A, 2021GeCoA.296...18J}.
In future work, it is necessary to address this issue by developing a model that takes into account the relative velocities among cloud particles.
Similarly, primary ejecta released from the planetesimal surface may undergo coalescence (or fragmentation) through mutual collisions before colliding with incoming debris dust or cloud particles \citep[e.g.,][]{2015Icar..250..215K}.
This issue also remains to be investigated in future studies.

The primary ejecta are eventually converted into cloud particles through momentum exchange during multiple collisions with incoming dust particles and cloud particles.
The stopping timescale of the primary ejecta, $t_{\rm stop}$, is then given by
\begin{equation}
t_{\rm stop} = \frac{m_{\rm ej} {( v_{\rm cl} - v_{\rm ej} )}}{A_{\rm ej} {\left[ \rho_{\rm d} {( v_{\rm pl} - v_{\rm ej} )}^{2} + \rho_{\rm cl} {( v_{\rm cl} - v_{\rm ej} )}^{2} \right]}},
\end{equation}
where $m_{\rm ej} = {( 4 \pi / 3 )} \rho_{\rm m} {R_{\rm ej}}^{3}$ is the mass of an individual primary-ejecta particle, and $A_{\rm ej} = \pi {R_{\rm ej}}^{2}$ is its collisional cross section.
The cloud thickness, $h_{\rm cl}$, is approximately given by
\begin{equation}
h_{\rm cl} = |v_{\rm ej}| t_{\rm stop},
\end{equation}
where $v_{\rm ej}$ is the mean ejecta velocity.
We define the positive velocity direction as that of the incoming debris dust; that is, $v_{\rm ej}$ is negative.

A fraction of the incoming dust particles reaches the planetesimal's surface, whereas the remainder is converted into cloud particles through collisions with primary ejecta or cloud particles.
The collision frequency of an incoming dust particle with primary ejecta is given by ${t_{\rm ej}}^{-1} = \rho_{\rm ej} A_{\rm ej} {\left( v_{\rm pl} - v_{\rm ej} \right)} / m_{\rm ej}$.
Similarly, the collision frequency of an incoming dust particle with cloud particles is given by ${t_{\rm cl}}^{-1} = \rho_{\rm cl} A_{\rm cl} {\left( v_{\rm pl} - v_{\rm cl} \right)} / m_{\rm cl}$, where $m_{\rm cl} = {(4 \pi / 3 )} \rho_{\rm m} {R_{\rm d}}^{3}$ is the mass of a cloud particle, and $A_{\rm cl} = \pi {R_{\rm d}}^{2}$ is its collisional cross section.
The timescale for incoming dust particles to traverse the dust cloud is $h_{\rm cl} / v_{\rm pl}$.
We then evaluate the fraction of incoming debris dust that reaches and collides with the planetesimal surface, $f_{\rm pl}$, as
\begin{equation}
f_{\rm pl} = \exp{\left[ - \frac{h_{\rm cl}}{v_{\rm pl}} \left( {t_{\rm ej}}^{-1} + {t_{\rm cl}}^{-1} \right) \right]}.
\end{equation}

At the planetesimal surface, colliding debris dust generates primary ejecta.
The ejecta particles are assumed to follow a power-law velocity distribution with a minimum ejecta velocity of $v_{\rm min}$ \citep[e.g.,][]{2011Icar..211..856H}.
In this study, we adopt $v_{\rm min} = 10~{\rm m}~{\rm s}^{-1}$ for the fiducial case shown in Figure~\ref{fig:6}.
We also examine the effect of larger $v_{\rm min}$ on the dust-cloud structure (see Figure~\ref{fig:e2}).
The mass flux ratio of ejecta to colliding debris dust, $x_{\rm in \to ej}$, is calculated from the crater-ejecta scaling law \citep[e.g.,][]{2011Icar..211..856H} as
\begin{equation}
x_{\rm in \to ej} = {\left( \frac{3 k}{4 \pi} \right)} {\left( \frac{v_{\rm pl}}{v_{\rm min}} \right)}^{3 \mu},
\end{equation}
where $k = 0.39$ and $\mu = 0.55$ are experimentally determined material constants for melt and rock \citep{2011Icar..211..856H}.
The mean ejecta velocity, $v_{\rm ej}$, is proportional to $v_{\rm min}$ and is given by
\begin{equation}
v_{\rm ej} = - {\left( \frac{3 \mu}{3 \mu - 1} \right)} v_{\rm min}.
\end{equation}
Similarly, fallback cloud particles colliding with the planetesimal surface generate primary ejecta.
The mass flux ratio of ejecta to colliding fallback cloud particles, $x_{\rm cl \to ej}$, is given by
\begin{equation}
x_{\rm cl \to ej} = {\left( \frac{3 k}{4 \pi} \right)} {\left( \frac{v_{\rm cl}}{v_{\rm min}} \right)}^{3 \mu}
\end{equation}
for $v_{\rm cl} > 0$, whereas $x_{\rm cl \to ej} = 0$ otherwise.

The mass and momentum conversion from the {\bf in} to the {\bf ej} components are caused by collisions of the {\bf in} component with the planetesimal surface.
Thus, $\dot{\Sigma}_{{\rm in} \to {\rm ej}}$ and $\dot{p}_{{\rm in} \to {\rm ej}}$ are given by
\begin{eqnarray}
\dot{\Sigma}_{{\rm in} \to {\rm ej}} & = & f_{\rm pl} x_{\rm in \to ej} \rho_{\rm d} v_{\rm pl} \frac{S_{\rm pl}}{S_{\rm cl}}, \\
\dot{p}_{{\rm in} \to {\rm ej}}      & = & \dot{\Sigma}_{{\rm in} \to {\rm ej}} v_{\rm ej}.
\end{eqnarray}
In contrast, the mass and momentum conversion from the {\bf in} to the {\bf cl} components are caused by collisions of the {\bf in} component with the {\bf ej} or {\bf cl} components before reaching the planetesimal surface.
Thus, $\dot{\Sigma}_{{\rm in} \to {\rm cl}}$ and $\dot{p}_{{\rm in} \to {\rm cl}}$ are given by
\begin{eqnarray}
\dot{\Sigma}_{{\rm in} \to {\rm cl}} & = & {\left( 1 - f_{\rm pl} \right)} \rho_{\rm d} v_{\rm pl}, \\
\dot{p}_{{\rm in} \to {\rm cl}}      & = & \dot{\Sigma}_{{\rm in} \to {\rm cl}} v_{\rm pl}.
\end{eqnarray}

The mass and momentum conversion from the {\bf cl} to the {\bf ej} components is caused by fallback of the {\bf cl} component onto the planetesimal surface.
Thus, $\dot{\Sigma}_{{\rm cl} \to {\rm ej}}$ and $\dot{p}_{{\rm cl} \to {\rm ej}}$ are given by
\begin{eqnarray}
\dot{\Sigma}_{{\rm cl} \to {\rm ej}} & = & x_{\rm cl \to ej} \rho_{\rm cl} v_{\rm cl} \frac{S_{\rm pl}}{S_{\rm cl}}, \\
\dot{p}_{{\rm cl} \to {\rm ej}}      & = & \dot{\Sigma}_{{\rm cl} \to {\rm ej}} v_{\rm ej}
\end{eqnarray}
for $v_{\rm cl} > 0$, whereas $\dot{\Sigma}_{{\rm cl} \to {\rm ej}} = 0$ and $\dot{p}_{{\rm cl} \to {\rm ej}} = 0$ otherwise.

The mass and momentum conversion from the {\bf ej} to the {\bf cl} components is caused by collisions of the {\bf ej} component with the {\bf in} and {\bf cl} components.
Thus, $\dot{\Sigma}_{{\rm ej} \to {\rm cl}}$ and $\dot{p}_{{\rm ej} \to {\rm cl}}$ are given by
\begin{eqnarray}
\dot{\Sigma}_{{\rm ej} \to {\rm cl}} & = & \frac{\Sigma_{\rm ej}}{t_{\rm stop}}, \\
\dot{p}_{{\rm ej} \to {\rm cl}}      & = & \frac{p_{\rm ej}}{t_{\rm stop}}.
\end{eqnarray}

In our model, cloud particles either accrete onto the planetesimal or leak sideways from the dust cloud as molten chondrules.
The removal rates from the dust-cloud system, $\dot{\Sigma}_{{\rm cl} \to {\rm ex}}$ and $\dot{p}_{{\rm cl} \to {\rm ex}}$, are given by
\begin{eqnarray}
\dot{\Sigma}_{{\rm cl} \to {\rm ex}} & = & \rho_{\rm cl} \cdot \max{\left( v_{\rm cl}, 0 \right)}, \\
\dot{p}_{{\rm cl} \to {\rm ex}}      & = & \dot{\Sigma}_{{\rm cl} \to {\rm ex}} v_{\rm cl}.
\end{eqnarray}

\section{Dependence on the ejecta radius}
\label{app:rad}

\begin{figure*}[]
\centering
\includegraphics[width = 0.7\textwidth]{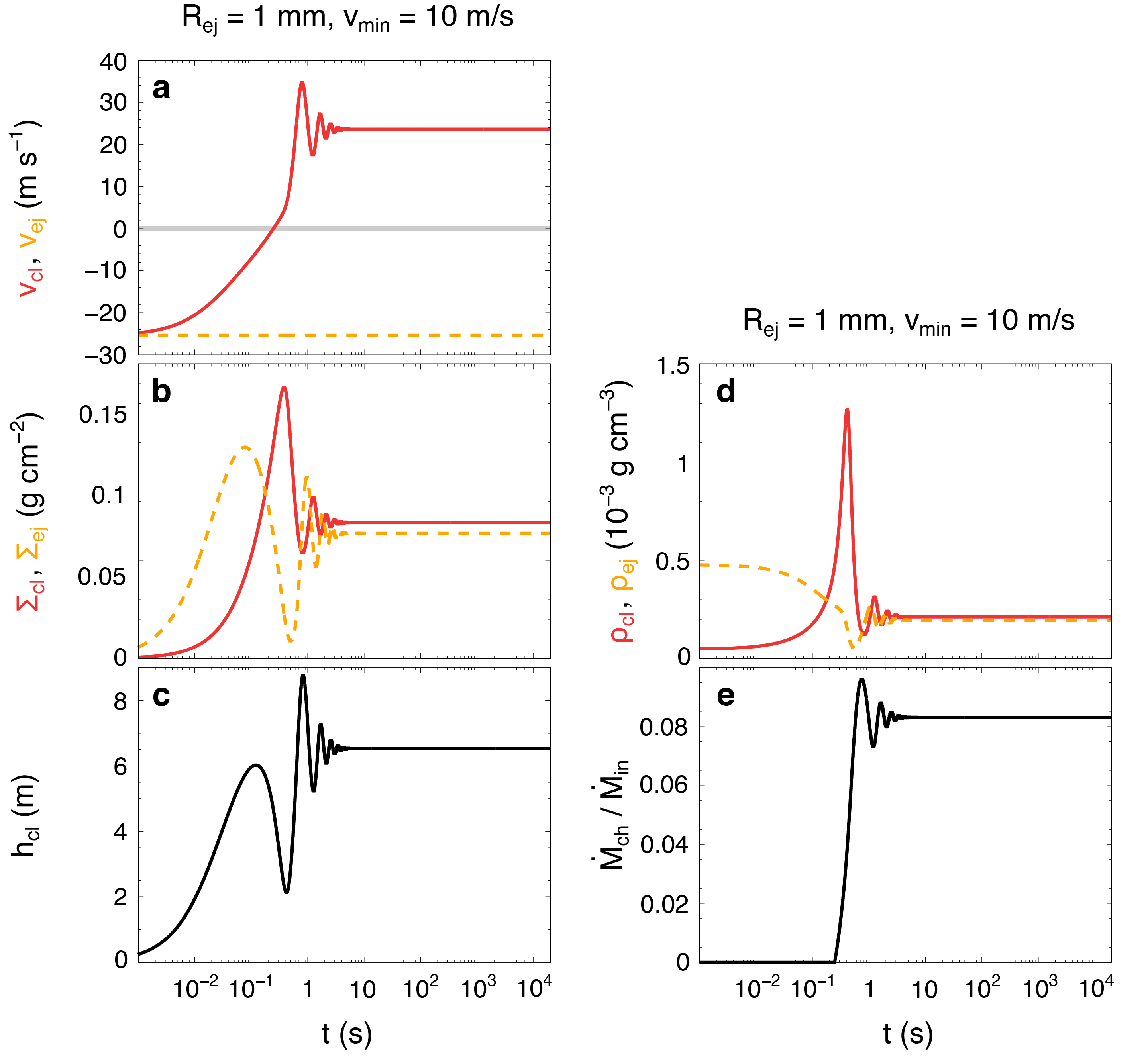}
\caption{
Same as Figures~\ref{fig:6}{\bf a}--\ref{fig:6}{\bf e}, but for $R_{\rm ej} = 1~{\rm mm}$ instead of $5~{\rm mm}$.
}
\label{fig:e1}
\end{figure*}

In this study, we adopt $R_{\rm ej} = 5~{\rm mm}$ as the fiducial value of the ejecta radius.
However, this choice is not constrained, and in practice $R_{\rm ej}$ should be determined by future detailed experiments.
The cloud thickness, $h_{\rm cl}$, is proportional to the stopping time $t_{\rm stop}$, which in turn scales with $m_{\rm ej} / A_{\rm ej}$, that is, with $R_{\rm ej}$.
Therefore, adopting a smaller $R_{\rm ej}$ is expected to lead to a smaller cloud thickness.

Figure~\ref{fig:e1} shows the structural evolution of the dust cloud system for the case with $R_{\rm ej} = 1~{\rm mm}$ instead of $R_{\rm ej} = 5~{\rm mm}$.
As discussed above, the smaller value of $R_{\rm ej} = 1~{\rm mm}$ results in a smaller steady-state $h_{\rm cl}$ than in the fiducial case (Figure~\ref{fig:e1}{\bf c}).
Consequently, the steady-state value of $\dot{M}_{\rm ch} / \dot{M}_{\rm in}$ is also smaller (Figure~\ref{fig:e1}{\bf e}).
If $\dot{M}_{\rm ch} / \dot{M}_{\rm in} \ll 1$, the amount of chondrules resupplied to the debris disk should be small.
It remains to be discussed whether this is consistent with observational evidence that meteoritic chondrules experienced multiple heating events, as well as with the total chondrule mass budget.

\section{Splashing criterion and minimum ejecta velocity}
\label{app:v_ej}

\begin{figure*}[]
\centering
\includegraphics[width = 0.7\textwidth]{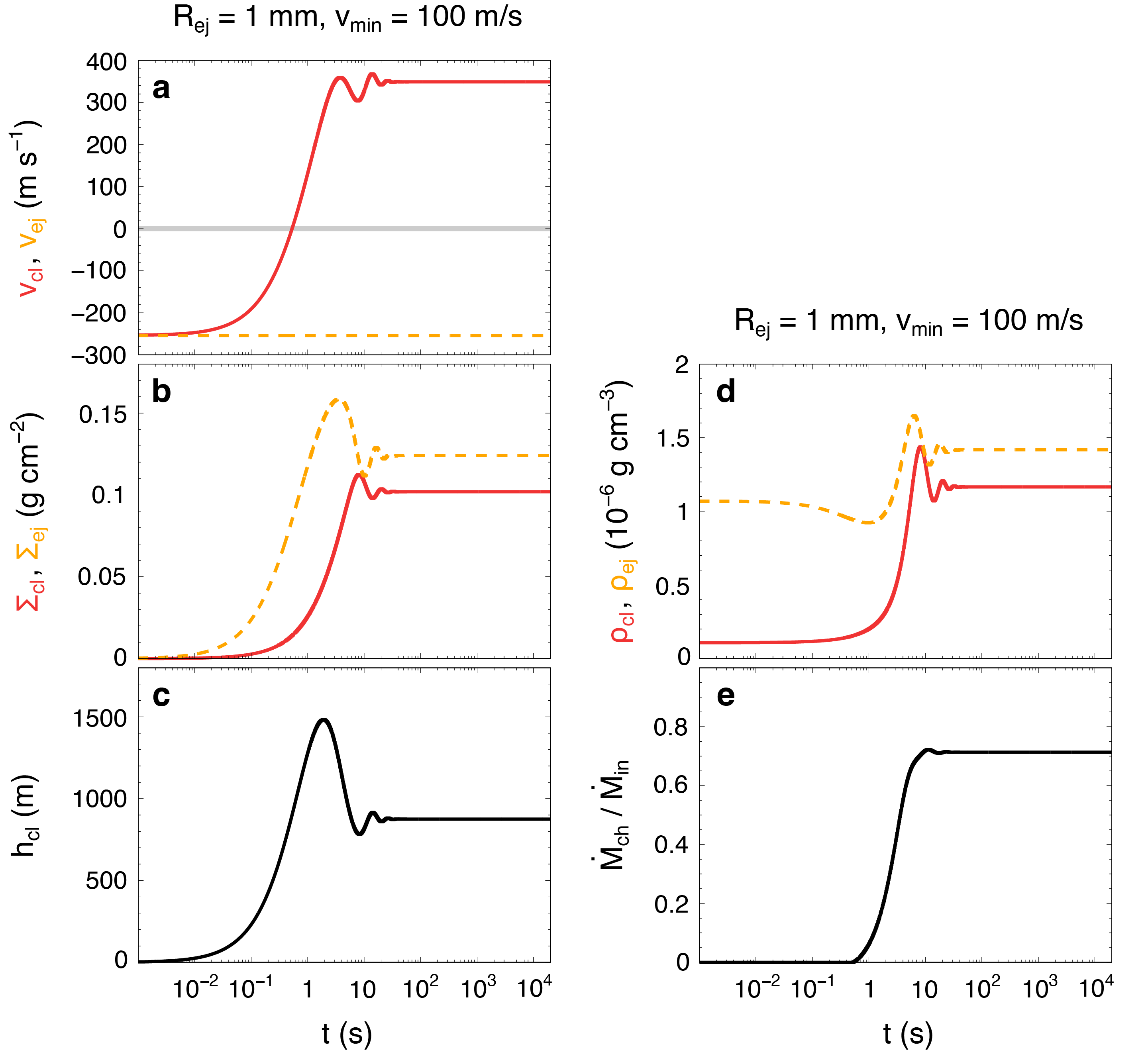}
\caption{
Same as Figures~\ref{fig:e1}{\bf a}--\ref{fig:e1}{\bf e}, but for $v_{\rm min} = 100~{\rm m}~{\rm s}^{-1}$ instead of $10~{\rm m}~{\rm s}^{-1}$.
}
\label{fig:e2}
\end{figure*}

Molten silicate droplets have high viscosity and surface tension, so low-speed collisions with the planetesimal surface are more likely to result in sticking than in splashing.
In particular, because fallback of the {\bf cl} component occurs at lower velocities than that of the {\bf in} component, it may not generate ejecta under some conditions.

If surface tension controls the energy dissipation, the critical condition can be described by the Weber number, ${\rm We}$.
Here, ${\rm We} \equiv 2 \rho_{\rm m} R_{\rm d} v_{\rm cl}^{2} / \sigma$, where $\sigma = 0.4~{\rm J}~{\rm m}^{-2}$ is the surface tension of silicate melt.
Splashing is expected to occur when collisions take place at ${\rm We} \gtrsim 10$ \citep[e.g.,][]{1990JFM...221..183A, 2005Icar..173..295K, 2008Icar..197..621K}, which corresponds to $v_{\rm cl} \gtrsim 3~{\rm m}~{\rm s}^{-1}$.
In contrast, if viscosity controls the energy dissipation, the critical condition can be described by the Reynolds number, ${\rm Re}$.
Here, ${\rm Re} \equiv 2 \rho_{\rm m} R_{\rm d} v_{\rm cl} / \eta$, where $\eta$ is the viscosity.
The viscosity of silicate melt depends strongly on its composition and temperature.
Splashing is expected to occur when collisions take place at ${\rm Re} \gtrsim 100$ \citep[e.g.,][]{2016ExFl...57..187S, 2019ApJ...877...84A}, which corresponds to $v_{\rm cl} \gtrsim 200~{(\eta / 1~{\rm Pa}~{\rm s})}~{\rm m}~{\rm s}^{-1}$.

From the above discussion, if the effect of viscosity is negligible, that is, if the impact outcome is controlled primarily by surface energy, we expect ejecta particles to be efficiently generated at the planetesimal surface when $v_{\rm cl} > 10~{\rm m}~{\rm s}^{-1}$.
Indeed, our numerical results satisfy this condition in the steady state for $t > 10~{\rm s}$ (Figure~\ref{fig:6}{\bf a}).
We also imagine that the minimum ejecta velocity, $v_{\rm min}$, is of the same order as the critical velocity for the sticking-to-splashing transition, and we therefore adopt $v_{\rm min} = 10~{\rm m}~{\rm s}^{-1}$.

We note that, if viscous effects are important, ejecta particles may not be generated by fallback of cloud particles unless $v_{\rm cl} \gg 10~{\rm m}~{\rm s}^{-1}$.
In this case, $v_{\rm min}$ would also be higher than the fiducial value of $10~{\rm m}~{\rm s}^{-1}$, and we therefore adopt $v_{\rm min} = 100~{\rm m}~{\rm s}^{-1}$ instead of $10~{\rm m}~{\rm s}^{-1}$.
Figure~\ref{fig:e2} shows the structural evolution of the dust cloud system for the case with $R_{\rm ej} = 1~{\rm mm}$ and $v_{\rm min} = 100~{\rm m}~{\rm s}^{-1}$.
The cloud thickness $h_{\rm cl}$ is proportional to $| v_{\rm ej} |$, which in turn scales with $v_{\rm min}$.
The larger value of $v_{\rm min} = 100~{\rm m}~{\rm s}^{-1}$ results in a larger steady-state value of $h_{\rm cl}$ than in the case with $v_{\rm min} = 10~{\rm m}~{\rm s}^{-1}$ (Figure~\ref{fig:e2}{\bf c}).
Consequently, the steady-state value of $\dot{M}_{\rm ch} / \dot{M}_{\rm in}$ is also larger (Figure~\ref{fig:e2}{\bf e}).

As discussed above, the conversion efficiency $\dot{M}_{\rm ch} / \dot{M}_{\rm in}$ varies significantly depending on the assumed value of $v_{\rm min}$.
Both laboratory experiments and theoretical studies to constrain $v_{\rm min}$ under realistic impact conditions will be important in future work.

\section{Peak temperature of planetesimals}
\label{app:T_peak}

Planetesimals that formed during the earliest stages of Solar System formation were heated internally by the decay energy of the short-lived radionuclide $^{26}$Al.
If the peak temperature at the center of a planetesimal, $T_{\rm peak}$, exceeds $\approx 1500~{\rm K}$, its interior may differentiate into silicate and metal components.
Here, we estimate $T_{\rm peak}$ for planetesimals with radii $R_{\rm pl} \lesssim 10~{\rm km}$, which are particularly important in our chondrule formation model.

The heating rate per unit mass due to the decay of $^{26}$Al can be written as $Q = Q_{0} \exp{( - t / \tau_{\rm d} )}$, where $t$ is the time after CAI formation and $\tau_{\rm d} = 1~{\rm Myr}$ is the mean lifetime of $^{26}$Al.
Here, $Q_{0}$ is the heating rate at $t = 0$.
For anhydrous chondritic planetesimals, $Q_{0} \approx 7~{\rm J}~{\rm kg}^{-1}~{\rm yr}^{-1}$ \citep[e.g.,][]{2012M&PS...47.2170S}, assuming that $^{26}$Al was spatially homogeneous throughout the protosolar disk and that its initial abundance was $^{26}{\rm Al}/^{27}{\rm Al} \approx 5 \times 10^{-5}$
\citep[e.g.,][]{2023Icar..40215607D}.

The thermal evolution of a planetesimal can be obtained by solving the heat diffusion equation.
Here, we consider the peak temperature in the limits where the planetesimal radius $R_{\rm pl}$ is much smaller or much larger than a critical radius $R_{\rm cr}$.
The critical radius is given by $R_{\rm cr} = \pi \sqrt{k \tau_{\rm d} / {( \rho c )}} \approx 10~{\rm km}$, where $k = 1~{\rm W}~{\rm m}^{-1}~{\rm K}^{-1}$, $\rho = 3000~{\rm kg}~{\rm m}^{-3}$, and $c = 1000~{\rm J}~{\rm kg}^{-1}~{\rm K}^{-1}$ are typical values of the thermal conductivity, density, and specific heat capacity of rocky materials, respectively \citep[e.g.,][]{1959chs..book.....C}.

When $R_{\rm pl} \ll R_{\rm cr}$, the peak temperature is determined by the balance between heating by the decay of $^{26}$Al and heat diffusion within the planetesimal.
It is approximately given by \citep[e.g.,][]{1959chs..book.....C}
\begin{equation}
T_{\rm peak} \approx T_{\rm s} + \frac{\rho Q_{0} }{6 k} {R_{\rm pl}}^{2} \exp{\left( - \frac{t}{\tau_{\rm d}} \right)},
\label{eq:T_peak}
\end{equation}
where $T_{\rm s}$ is the surface temperature of the planetesimal.
For example, substituting $t = 1~{\rm Myr}$, $R_{\rm pl} = 2~{\rm km}$, and $T_{\rm s} = 300~{\rm K}$ into Equation~\eqref{eq:T_peak} gives $T_{\rm peak} \sim 500~{\rm K}$.
This temperature is far below $1500~{\rm K}$, indicating that such a small planetesimal is unlikely to undergo silicate--metal differentiation.

In contrast, for $R_{\rm pl} \gg R_{\rm cr}$, the peak temperature is determined by the time-integrated heating due to the decay of $^{26}$Al.
In this limit, $T_{\rm peak}$ is approximately given by $T_{\rm peak} \approx T_{\rm s} + {( Q_{0} \tau_{\rm d} / c )} \exp{( - t / \tau_{\rm d} )}$.
Using $t = 1~{\rm Myr}$ and $T_{\rm s} = 300~{\rm K}$, we obtain $T_{\rm peak} \sim 3000~{\rm K}$.
This temperature is far above $1500~{\rm K}$, indicating that such an early-formed, large planetesimal is likely to undergo silicate--metal differentiation and may correspond to an achondrite parent body.


\bibliography{sn-bibliography}{}
\bibliographystyle{aasjournalv7}



\end{document}